\documentclass[pra,letterpaper,onecolumn]{revtex4-1}   % for review and submission
\usepackage{amsbsy,amssymb,amsmath,bm,bbold} 
\usepackage{graphicx,color,epsfig,rotate} 
\usepackage{fancyhdr} 
\usepackage{epstopdf}
\usepackage{csquotes}
\usepackage{float}
\usepackage[colorlinks=true,linkcolor=blue,citecolor=blue,urlcolor=blue]{hyperref}
\usepackage{gensymb}
\usepackage{times,xspace}
\begin{document}
	
	% The following information is for internal review, please remove them for submission

	% the following line is for submission, including submission to the arXiv!!
	%\hspace{5.2in} \mbox{Fermilab-Pub-04/xxx-E}
	
	\title{ Momentum controlled optical pulse amplification in photonic time crystals}
	
	\author{Snehashis Sadhukhan,$^1$ and Somnath Ghosh$^{1,}$}
	\email{somiit@rediffmail.com}
	\affiliation{\vspace{0.3cm}
		$^1$Department of Physics, Indian Institute of Technology Jodhpur, Rajasthan-342030, India}

	\begin{abstract}
We show that by manipulating the momentum ($k$) of a propagating optical pulse, the intensity can be controlled and highly enhanced in a linear binary photonic time crystal (PTC) system.
The optical pulse equipped with $k$ lying within the bandgap of the PTC gets amplified and gives rise to reflected and transmitted pulses with an equal growth in intensity moving in opposite directions.
We predict and quantify the amount of amplification of the transmitted and reflected pulse and show
the amplifications achieve a maximum at the centre of the $k$-gap accompanied by weaker growth in intensity at the edges. 
The maximum growth in the intensity of the optical pulses attains different values for different ranges of $k$-gaps in such systems. 
Such precise control over the amplification of propagating pulse can be exploited exclusively by tailoring the wave vector of the pulse and open a unique platform for light manipulation in futuristic unconventional active photonic devices.
	\end{abstract}

	\maketitle
	
	%\section{\label{sec:level1}First-level heading}
	% sections are not used for PRL papers
In wake of tremendous scientific and technological advancements in the last decade,  photonic time crystals (PTCs) have been introduced as a distinct tool for light manipulation with unique features opening up new avenues in the domain of photonics. 
PTCs are systems where the refractive index ($n$) varies periodically with time \cite{shaltout2016photonic, zeng2017photonic}, the study of which has brought light to many underlying features like temporal localization in between two PTCs with dissimilar topologies \cite{lustig2018topological}, localization of pulses and exponential decay in group velocity of a pulse in disordered PTCs \cite{sharabi2021disordered}, exceptional points in spatio-temporal crystals \cite{sharabi2022spatiotemporal}, and more with their intriguing physics. 
The concept of PTCs, so far, has been demonstrated in transmission lines for radio frequencies \cite{reyes2015observation} and tremendous efforts are being put forward towards their realisation in optical regime using transparent conducting oxides (TCOs) \cite{lustig2021towards}. 
Moreover, in PTCs, the periodic temporal variation of $n$ yields an exclusive dispersion diagram which is gaped in momentum ($k$) \cite{shaltout2016photonic,lustig2018topological, ma2019band} leading to a range of unconventional signatures like pulse amplification, pulse broadening etc. The amplification as of the currently available literature is an integral part of the gap propagation (when $k$ of the propagating pulse lies within the gap) in PTCs.
\\
The pulse amplification in PTCs is a consequence of broken continuous time translation symmetry \cite{zeng2017photonic}. 
Such kind of pulse amplification in PTCs is fundamentally different from optical parametric amplification (OPA) \cite{sharabi2022spatiotemporal}. 
When a pulse enters a PTC along with exponential growth in intensity, the group velocity of the propagating pulse decreases simultaneously and eventually becomes zero \cite{lustig2018topological, sharabi2021disordered}. Moreover, the pulse suffers from spatial and temporal broadening due to dispersion in the medium. An amplification of order four of magnitude is accompanied by a broadening of 2.5 times can be observed under such a scenario \cite{lustig2018topological}. 
When the PTC is over, the pulse emerges as two pulses propagating in opposite directions. \\
Time reflection and refraction of the propagating pulse in the optical regime require a sudden and abrupt periodic change in $n$ that leads to
formulation of significant $k$-gaps.
While achieving a gap propagation, the strong interference among these periodically reflected and transmitted pulses give rise to eigenmodes with imaginary Floquet frequencies which result in an amplification of the pulse \cite{sharabi2021disordered}. \\
The amplification in PTCs has huge potential to fuel state-of-the-art lasing technologies and other active photonic devices. 
To exercise control over such amplification it is important to characterize the amplification and its dependence.
However, the present literature lags behind the theoretical treatment that deals with pulse amplification within the k-gap of PTCs. 

In this work, we report for the first time, the $k$ dependent amplification of a propagating pulse within the $k$-gap of a linear binary PTC system. We investigate the pulse dynamics in side and out side of the bandgap and study the amplification in time-reflected and transmitted pulses. Further, we quantify the amount of amplification using our proposed analytical modelling and show that the growth in the intensity of the time-reflected and transmitted waves for a pulse entering a PTC having $k$ within the BG are of the same order. 
Beyond demonstrating  analytically, we establish the findings using finite difference time domain (FDTD) simulations that the amplification achieved are different at various ranges of $k$-gaps. Additionally, the study reveals that the amplification is non-uniform throughout a $k$-gap and attains a maximum at the centre of the gap. 
\begin{figure}[b!]
		\includegraphics[width=8.5cm]{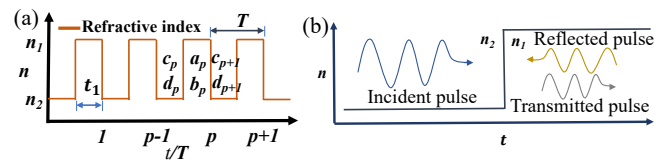}
		\caption{(a) Schematic of a binary photonic time crystal where permittivity varies periodically in a step-like manner with time. For $0<t<t_2$ seconds $n (t) =n_2$, and $t_2<t<T-t_2$ seconds $n (t) =n_1$ with a time period of $T$ (= $t_1+t_2$). (b) During time reflection at temporal discontinuity, reflected and transmitted waves in the same medium due to causality.}
		\label{fig1}
\end{figure}

%%%
%%% Results and Discussion	
%%% 

To unveil the underlining features of the amplification in a PTC, we first analytically calculate the band diagram of a PTC using the transfer matrix method that results in ranges of $k$-gaps where the PTC manifests profound amplification.
We begin by considering a plane wave propagating in a spatially homogeneous linear binary PTC [Fig. \ref{fig1}(a)] and describe the dynamics using Maxwell's equations for the electric displacement field $D(z,t)$ coupled to the Floquet-Bloch theorem.
The plane-polarised wave propagating in $z$ direction through such time-dynamic media undergoes time reflections at the temporal interfaces, and after the reflection, both time-reflected and time-refracted waves coexist within the same medium [Fig. \ref{fig1}(b)].
These time-reflected and transmitted waves interfere to produce the Floquet-Bloch modes in PTCs.
For simplicity, we consider a non-magnetic media.
To obtain significant $k$-gaps to influence the plane wave propagation through PTC, a refractive index difference of the order of unity is required and we choose the permittivity ($\epsilon$) in general  varying from $\epsilon_2$ ($=1)$ (for $t_2$ fs) to $\epsilon_1$ ($=3$) (for $t_1$ fs) in a periodic-step like fashion in time that results in a binary PTC [see Fig. \ref{fig1} (a)].
Since in a practical photonic system the time-reflection vanishes for changes slower than the $4.3$ fs time period \cite{lustig2021towards}, the binary variation is considered to observe the time reflection in  the analytical model.
Thus we choose a time period ($T$) of 2 fs and considered $t_1=t_2=1$ fs.
During the time reflections at temporal interfaces, the momentum is conserved but temporal frequency gets modified and the instantaneous $D_x(z,t)$ within the PTC is written as a sum of forward propagating and time-reflected waves as
\begin{widetext}
\begin{equation}
D_x(z,t)=
\begin{cases}
      (a_p e^{i\omega_1 (t-pT)} + b_p e^{-i\omega_1 (t-pT)} )e^{-ik_zz}, & pT-t_1<t<pT\\
      (c_p e^{i\omega_2 (t-pT+t_1)} + d_p e^{-i\omega_2 (t-pT+t_1)} )e^{-ik_zz}, & (p-1)T<t<pT-t_1 \\
    \end{cases} 
    \end{equation}
\end{widetext}
where, $p$ is the $p^{th}$ unit cell in time and the frequency of the pulse getting modified as $\omega_{1,2}=kc/n_{1,2}$ ($c$= velocity of light in free space) keeping the $k$ constant at the interfaces, and $a_p$, $b_p$, $c_p$ and $d_p$ are constants which are related by the continuity conditions at the temporal interfaces involving $D_x$ and $B_y$. 
Utilizing the continuity of $D_x$ and $B_y$ at $t= pT-t_1$ and $t= pT$ interfaces we obtain the matrix equation

\begin{figure}[b!]
		\includegraphics[width=8.5cm]{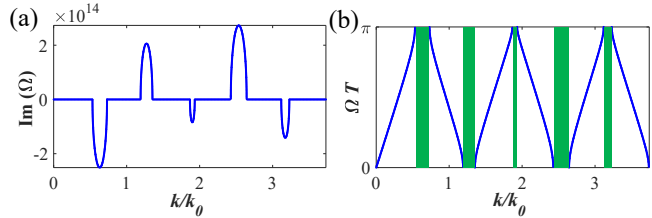}
		\caption{(a) Imaginary part of $\Omega$ as a function of normalized $k$ (where, $k_0=2\pi /Tc$). (b) Band diagram of a PTC with infinite number of bilayers in time, showing allowable bands (in blue) representing ranges of $k$ where $\Omega$ is real and bandgaps (BGs) (in green) where $\Omega$ is imaginary.}
		\label{fig2}
\end{figure}
\begin{gather}
 \begin{bmatrix} c_{p}  \\ d_{p}  \end{bmatrix}
 =  \begin{bmatrix}
   P  &
   Q \\
   R &
   S 
   \end{bmatrix}  \begin{bmatrix} c_{p+1}  \\ d_{p+1}  \end{bmatrix}
   \label{Transfer_matrix}
\end{gather}
where, 
\begin{subequations}
\begin{equation}
  P= e^{i\omega_2 t_2} (\cos(\omega_1 t_1) + \frac{i}{2} (\frac{\omega_1}{\omega_2}+\frac{\omega_2}{\omega_1}) \sin(\omega_1 t_1)
\end{equation}    
\begin{equation}
  Q=  \frac{i}{2} e^{i\omega_2 t_2} (\frac{\omega_1}{\omega_2}+\frac{\omega_2}{\omega_1}) \sin(\omega_1 t_1)
\end{equation}
\end{subequations}
and $S=P^*$, $R=Q^*$.
To get the dispersion relation using this matrix relation we recall the $D_x(z,t)$ for the time-periodic system exhibiting $n(t)=n(t+T)$ follow the Floquet-Bloch theorem and modifies as $D_{\Omega}(t)e^{-i\Omega t} e^{ikz}$. Thus we write
\begin{equation}
    \Omega(k_z)= \frac{1}{T} \cos^{-1} [\frac{P+S}{2} ].  
\end{equation}
By plotting $\Omega$ as a function of $k_z$ the band diagram for the given PTC [Fig. \ref{fig2}] with infinite number of temporal unit cells is obtained and the same is used to
recognize the BG ranges in $k$.
We obtain bands where $\Omega$ is real and bandgaps for $\Omega$ being complex [see Fig. \ref{fig2}(a)].
The first few (five in this case) of  such BGs ranges in $k$ are as follows $0.514$ to $0.731$ (which we term as BG1), $1.191$ to $1.353$ (BG2), $1.867$ to $1.933$ (BG3), $2.426$ to $2.644$ (BG4), $3.118$ to $3.229$ (BG5).
Moreover, we note the corresponding width of the BGs $0.217$, $0.162$, $0.066$, $0.218$, $0.111$ respectively (in terms of $k/k_0$).
From Fig. \ref{fig2} we observe that the wider BGs produce higher values of Im($\Omega$) within the BG. 
Moreover, the PTC amplifies  plane waves with $k$ lying within these BGs.
%%%%
%%%%   END: Band DIAGRAM
%%%%
%%%%
%%%% START: BAND PROPAGATION GAP PROPAGATION
%%%%

\begin{figure}[t!]
		\includegraphics[width=8.5cm]{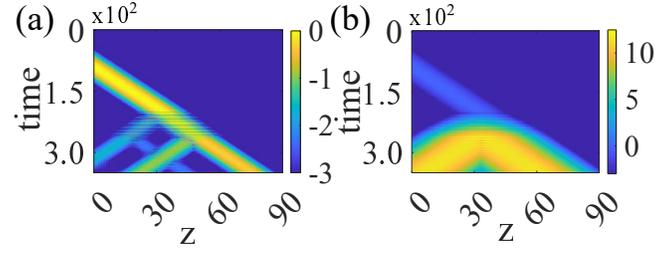}
		\caption{Pulse intensity profile for propagation through a PTC system depicting 
(a) pulse splitting at the beginning and end of PTC due to time reflections with no amplification in case of the band propagation, (b) pulse amplification in addition to localization (in space) and pulse broadening for gap propagation. Colour bars in log scale.
}
		\label{fig3}
\end{figure}

To establish and explore the fundamental characteristics of wave propagation through PTC we simulate the pulse propagation and analyze the effect of the $k$ gaps.
We simulate the PTC by solving Maxwell's equation using FDTD method by judiciously implementing Yee's grid method and following the stability criteria for PTCs \cite{zeng2017photonic, yee1966numerical}.
We consider a gaussian pulse with full width at half maxima (FWHM) of $50$ fs.
The PTC starts at $200$ fs, and ends at $270$ fs.
At first, the pulse propagates through the vacuum and it enters the system at $200$ fs.
We simulate pulse propagation for two particular cases one for the pulse with $k$ lying within the band $k= 0.4k_0$ (which is the band propagation) [Fig. \ref{fig3}(a)] and other one for the $k$ lying within the bandgap $k= 0.62k_0$ (gap propagation) [Fig. \ref{fig3}(b)].
As we can see from Fig. \ref{fig3}(a) for the first case the propagating pulse does not get amplified and we obtain two transmitted and two time-reflected pulses at the end of the PTC. Moreover, the maximum amount of energy is transferred to the transmitted pulse leaving a relatively small amount of energy into the time-reflected pulses. 
Whereas, for the gap propagation an exponential growth in the pulse intensity of $10^{14}$ is observed [Fig. \ref{fig3}(b)] and at the end of the PTC we are left with one transmitted and one time-reflected pulse, where both get equally amplified.
Here, being equipped with a $k$  not allowed to propagate through the PTC, pulse localization in space with spatial and temporal pulse broadening along with the amplification are also observed as a characteristic feature of gap propagation.
%%%%%
%%%% END
%%%%%
%%%%
%%%%   START: Transmission-Reflection
%%%%

\begin{figure}[t]
		\includegraphics[width=8.5cm]{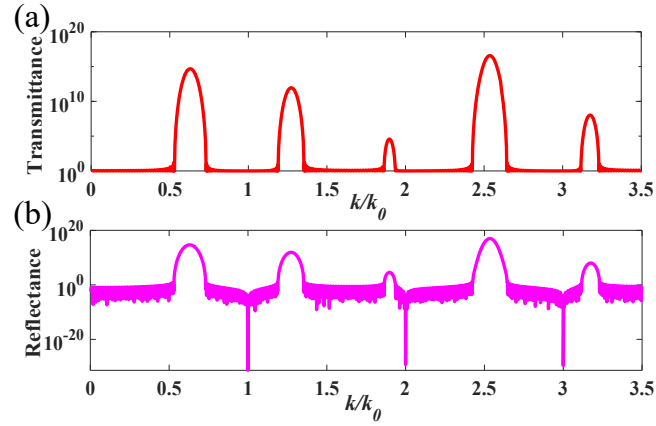}
		\caption{ (a) Transmittance and (b) reflectance plot of PTC showing amplification of the forward propagating and the time-reversed pulses within the BGs. The amplification is maximum at the middle of the BG and is different at different ranges of BG.
}
		\label{fig4}
\end{figure}

In view to predict and quantify the amount of amplification theoretically and 
have a better insight into the amplification phenomenon in PTC, the transmission and reflection of a plane wave are studied.
In a practical senario, the number of temporal unit cells ($N$) is always limited by the system definition and requires to be of a finite number.
As a result, for a plane wave propagating through $N$ number of temporal unit cells, the $D$ consists of two components one forward propagating ($a_p$ or $c_p$) and time reflected ($b_p$ or $d_p$) at the $p^{th}$ unit cell depending on the portion of temporal unit cell where the wave is in.
It is evident from the PTC simulation [see Fig. \ref{fig3}] that before entering the PTC irrespective of its $k$,  the $d$ component (time-reflected part) of the field just before entering the system ($d_0$) is zero for a PTC. 
In this scenario, the effect of $N$ temporal unit cells can be taken into account by simply multiplying
Eq. \ref{Transfer_matrix} $N$ times once for each  cell in time. Thus, we obtain 
\begin{gather}
 \begin{bmatrix} c_{0}  \\ 0  \end{bmatrix}
 =  \begin{bmatrix}
   P  &
   Q \\
   R &
   S 
   \end{bmatrix}^N   \begin{bmatrix} c_{N}  \\ d_{N}  \end{bmatrix}
   =\begin{bmatrix}
   A  &
   B \\
   C &
   D 
   \end{bmatrix}   \begin{bmatrix} c_{N}  \\ d_{N}  \end{bmatrix}
   \label{TM_2}
\end{gather}
Eq. \ref{TM_2} relates the coefficients of $D$ of the wave before and after the PTC completing $N$ temporal periods. 
Thus we get
\begin{subequations}
\begin{align}
 c_0 &= A c_N + B  d_N  \\
0 & = C c_N +D d_N
\end{align}
\label{subeq1}
\end{subequations}
Here we define transmission coefficient for the displacement field amplitude for PTC as 
\begin{equation}
    t_N=(\frac{c_N}{c_0})_{d_0 =0}
    \label{t_n}
\end{equation}
Since the time-reflected pulse is observed at the end of the PTC, the reflection coefficient for the displacement field amplitude turns out to be
\begin{equation}
    r_N=(\frac{d_N}{c_0})_{d_0 =0}
    \label{r_n}
\end{equation}
Using Eq. \ref{TM_2}- \ref{r_n}, the amplitude transmission and reflection coefficients can be rewritten as
\begin{subequations}
\begin{align}
  t_N &=\frac{1}{A-\frac{CB}{D}}  \\
r_N &=\frac{1}{B-\frac{AD}{C}} 
\end{align}
\label{t_n2} 
\end{subequations}
The transmitivity and reflectivity of the intensity i.e. the transmittance and reflectance is calculated by calculating the absolute value from Eq. \ref{t_n2} as
\begin{subequations}
\begin{align}
|t_N|^2 &=\frac{1}{|A-\frac{CB}{D}|^2}  \\
|r_N|^2 &=\frac{1}{|B-\frac{AD}{C}|^2} 
\end{align}
\label{abst_n2} 
\end{subequations}
We plot the $|t_N|^2$ and $|r_N|^2$ as a function of momentum $k$ in Fig. \ref{fig4} (a) and (b) respectively and obtain the output intensity for both the transmitted and time reflected waves in PTC, analytically.  
%%%
%%% Explaining Tx Rx plot fig.4
%%
In Fig. \ref{fig4} (a), we observe the first band (i.e. $0<k/k_0<0.514  $) and it is evident that the transmission coefficient remains almost one for $k$ values ranging within the band which indicates that for plane waves with $k$ within the band there is no amplification which is the case of band propagation and the same scenario is true for all other bands. 
A similar analysis of the reflection plot [Fig. \ref{fig4}(b)] for the bands  clearly shows that only a certain fraction of the total intensity is reflected back within the range.
As we move to the BG1 (in case of the gap propagation) the value of $t_N$ starts rising, indicating the pulse amplification within the $k$ gap.
The $t_N$ attains a maximum value at the center of the BG1. 
Moreover, as we start moving away from the BG centre, the $t_N$ value starts falling and again it reaches to unity at the end of the BG1.
The same observation is true for all the other BG ranges.
Thus, for a PTC, amplification within a single $k$ gap is not uniform.
The amplification takes a certain value depending on the choice of value of the momentum and it is the maximum for a $k$ value at the centre of the BG.
It is evident from the reflectance plot in Fig. \ref{fig4}(b) for the gap propagation that the time-reflected wave generated at the end of the PTC is also equally amplified along with the transmitted wave and show similar characteristics of attending a maxima at the gap center.
Moreover, both the plots Fig. \ref{fig4} (a) and (b) show that the amplification obtained (in transmitted and time-reflected wave) at different the BG ranges is not the same and it is different for different ranges of BGs. 
Additionally, we observe different BG centers are producing different amount of amplifications. 
As from our analytical results, the estimated amplification at the BG centre for BGs are $\sim 10^{14}$ (BG1), $\sim 10^{11}$ (BG2), $\sim 10^{4}$ (BG3), $\sim 10^{16}$ (BG4), $\sim 10^{8}$ (BG5) respectively.

\begin{figure}[b]
		\includegraphics[width=7.5cm]{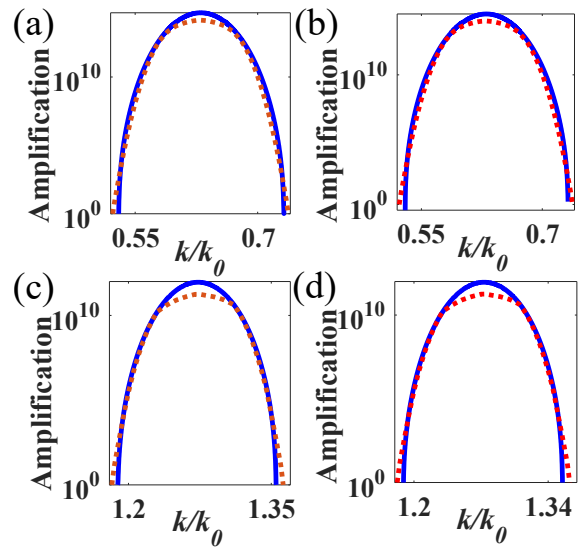}
		\caption{Comparison of the amplification of the (a, c) transmitted and (b, d) time-reflected pulse within the BG1 (a, b) and BG2 (c, d) (red dash: simulated data, blue line: analytical results).
}
		\label{fig5}
\end{figure}

%%%%%%%
%%%%%%% START: simulation for verification
%%%%%%%

\begin{figure}[t]
		\includegraphics[width=7.5cm]{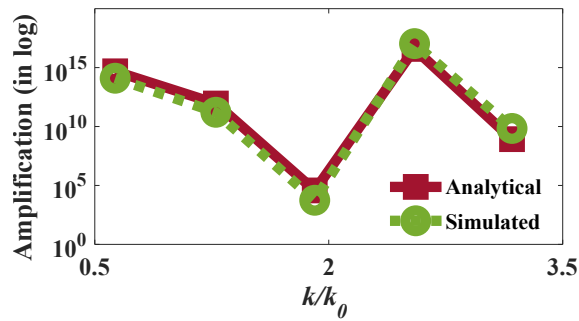}
		\caption{Amplification of the optical pulse at several BG centers (in red) and comparison with simulated data (in green).
}
		\label{fig6}
\end{figure}

To verify these properties of PTCs and validate our theoretical findings, we simulated a gaussian pulse propagation through a PTC.
This time we keep all the parameters same as in the previous case [Fig. \ref{fig3}] but select the $k$ of the pulse.
We observe the amplification of the transmitted and time-reflected pulse for BG1 and BG2 and plot them against $k$.
Observing the amplification from simulation data lead us to Fig. \ref{fig5}, where the amplification of the time reflected and transmitted pulse has been plotted and compared with the same derived from the analytical model.
For the BG1, the transmitted and time-reflected pluses are amplified in the order of $\sim 10^{14} $ in both cases [See Fig. \ref{fig5}]. Whereas, for the BG2, both pluses are amplified in the order of $\sim 10^{11} $ [See Fig. \ref{fig5}]. 
Thus, the simulated results are in complete agreement with the proposed analytical model and both of them show a growth in intensity of the same order.
Moreover, Fig. \ref{fig5} also verifies the fact that the amplification attains a maxima at the center of the BGs.
%%%%%
%%%%% WHY there is differece between FDTD and ANalytical
A careful observation of the Fig. \ref{fig5} reveals that the simulated curves are a bit widen than the analytical one and there exists a slight difference in amplifications between them.
Such difference attributes to two different schemes.
In case of analytical calculations, plane wave has been considered whereas in simulations pulse propagation is utilized.
With judiciously chosen grid resolutions and pulse parameters, the difference can be minimised to a value as low as $<5\%$.
%%%%
%%%% START: FIG 6 explain
Further to verify that the different BGs produce different amount of amplifications, we simulate the pulse propagation with $k$ values at the gap centers i.e. $0.63$ (BG1), $1.27$ (BG2), $1.91$ (BG3), $2.55$ (BG4), $3.17$ (BG5).
We compare the simulated results with the analytical one [in Fig. \ref{fig6}]. 
As from our simulated results, the amplification at the BG centre for BGs are $\sim 10^{14}$ (BG1), $\sim 10^{11}$ (BG2), $\sim 10^{4}$ (BG3), $\sim 10^{17}$ (BG4), $\sim 10^{9}$ (BG5), respectively which are in agreement with the analytically calculated values and verifies our observations from analytical calculations.
 
%%%%%%%%%%EXPLAINATION
%%%%%%%%%%
Amplification in the gap propagation in PTCs is observed due to their periodic variation in time that break continuous time translation symmetry,
as a result does not conserve energy. 
In gap propagation, the $\Omega$ becomes imaginary and as a result the forward propagating and time-reflected waves generated periodically within the PTC interfere to generate an envelope function of the form $\sim \sinh \Omega t + \cosh \Omega t$.
By comparing Figs. \ref{fig2}(a) and (b), we note that wider bandgaps result in larger values of Im($\Omega$) within the BG as well as at the gap center and larger values of Im($\Omega$) generate higher amplification within the BGs.
The BG in Fig. \ref{fig2} has been derived by considering infinite number of temporal unit cells ($N$). 
By limiting the $N$ for practical applications, we limit the amount of amplification with increasing number of $N$ the amplification asymptotically approaches the theoretically predicted value in as case of infinite $N$.
In PTCs the $n$ varies periodically in time and this variation is generated by a external agency which is the driver system.
Now $\Omega$ is a function of the driving period ($T$) which controls the BGs.
Thus, the extra energy is contributed by the driver system.

%%%%%%%%%%%%%
%%%%CONCLUSION
%%

In conclusion, we have shown that the amplification of the propagating optical pulse depends on the momentum within the BG of PTC, and the amplification is maximum at the center of a specific BG.
Moreover, the analytical modeling successfully quantifies the amount of amplification of the transmitted and time-reflected pulses propagating in opposite directions and establish the fact that different BGs produce different amount of amplifications.
Such new observations, involving engineering of the optical properties in PTCs, introduce a new degree of freedom to control the light and may open up a new avenue for future unconventional lasing and other active device applications in photonics.

%%%%%
%%%%%
%%%%%

	\bibliography{ss_ref}
	
\end{document}